# New copulas based on general partitions-of-unity and their applications to risk management (part II)

Dietmar Pfeifer[1], Andreas Mändle[1], and Olena Ragulina[2]

Carl von Ossietzky Universität Oldenburg, Germany[1] and
Taras Shevchenko National University of Kyiv, Ukraine[2]

October 23, 2017

**Abstract:** We present a constructive and self-contained approach to data driven infinite partition-of-unity copulas that were recently introduced in the literature. In particular, we consider negative binomial and Poisson copulas and present a solution to the problem of fitting such copulas to highly asymmetric data in arbitrary dimensions.

## 1. Introduction

Infinite partition-of-unity copulas have been introduced recently in the paper by *Pfeifer et al.* (2016). The main emphasis there was, however, on a particular symmetric case called diagonal dominance for which tail dependence coefficients could be explicitly calculated. The general asymmetric case was not treated in full detail. Our particular interest here is to complete the general setup with a suggestion how a data driven approach can be used to fit such copulas to highly asymmetric data in arbitrary dimensions, a question that had remained open so far.

## 2. A formal framework for infinite partition-of-unity copulas

Assume that $\{\varphi_{ki}(u)\}_{i \in \mathbb{Z}^+}$ for $k = 1, \cdots, d \in \mathbb{N}$ represent discrete distributions over $\mathbb{Z}^+$ with a parameter $u \in (0,1)$, i.e.

$$\varphi_{ki}(u) \geq 0 \quad \text{and} \quad \sum_{i=0}^{\infty} \varphi_{ki}(u) = 1 \text{ for } u \in (0,1),$$

with

$$\alpha_{ki} := \int_0^1 \varphi_{ki}(u)\,du > 0 \text{ for } i \in \mathbb{Z}^+.$$

Let further $\{p_\mathbf{i}\}_{\mathbf{i} \in \mathbb{Z}^{+d}}$ represent the distribution of an arbitrary discrete $d$-dimensional random vector $\mathbf{Z}$ over $\mathbb{Z}^{+d}$ where, for simplicity, we write $\mathbf{i} = (i_1, \cdots, i_d)$, and

$$P(\mathbf{Z} = \mathbf{i}) = p_\mathbf{i}, \ \mathbf{i} \in \mathbb{Z}^{+d}.$$

Suppose further that for the marginal distributions, there holds

$$P(Z_k = i) = \alpha_{ki}, \ i \in \mathbb{Z}^+, \ k = 1, \cdots, d.$$

Then

$$c(\mathbf{u}) := \sum_{\mathbf{i} \in \mathbb{Z}^{+d}} \frac{p_{\mathbf{i}}}{\prod_{k=1}^{d} \alpha_{k,i_k}} \prod_{k=1}^{d} \varphi_{k,i_k}(u_k), \quad \mathbf{u} = (u_1, \cdots, u_d) \in (0,1)^d \qquad (1)$$

defines the density of a *d*-variate copula, which is called *infinite partition-of-unity copula* (IPU-copula for short).

Alternatively, we can rewrite (1) as

$$c(\mathbf{u}) = \sum_{\mathbf{i} \in \mathbb{Z}^{+d}} p_{\mathbf{i}} \prod_{k=1}^{d} f_{k,i_k}(u_k), \quad \mathbf{u} = (u_1, \cdots, u_d) \in (0,1)^d \qquad (2)$$

where the $f_{ki}(\bullet) = \dfrac{\varphi_{ki}(\bullet)}{\alpha_{ki}}$, $i \in \mathbb{Z}^+$, $k = 1, \cdots, d$ denote the Lebesgue densities induced by the $\{\varphi_{ki}(u)\}_{i \in \mathbb{Z}^+}$. A stochastic representation of the probability distribution induced by (1) or (2) is given by the random vector $\mathbf{U}(\mathbf{Z}) = \mathbf{U}(Z_1, \cdots, Z_d) := (U_{1,Z_1}, \cdots, U_{d,Z_d})$ (with $\mathbf{Z}$ as above) with stochastically independent random variables $\{U_{ki} | k = 1, \cdots, d, i \in \mathbb{Z}^+\}$ (also independent of $\mathbf{Z}$) where the distribution of $U_{ki}$ is induced by the density $f_{ki}(\bullet) = \dfrac{\varphi_{ki}(\bullet)}{\alpha_{ki}}$, $i \in \mathbb{Z}^+$, $k = 1, \cdots, d$. $\mathbf{Z}$ is called the *driver* of the IPU copula with density given in (1).

The following two classes of IPU copulas have been investigated in detail in *Pfeifer et al.* (2016), among others:

**Example 1** (negative binomial copula): Let, for fixed integers $a_k > 0$,

$$\varphi_{ki}^{NB}(u) = \binom{a_k + i - 1}{i} u^i (1-u)^{a_k} \quad \text{for } k, i \in \mathbb{Z}^+ \text{ and } u \in (0,1). \qquad (3)$$

Here we have $\alpha_{ki}^{NB} = \int_0^1 \varphi_{ki}^{NB}(u)\,du = \dfrac{a_k}{(a_k + i)(a_k + i + 1)}$ which corresponds to a discrete analogy of a Pareto distribution. The densities $f_{ki}^{NB}$ are those of a beta distribution with parameters $(i+1, a_k+1)$. The corresponding copula density is thus given by

$$c^{NB}(\mathbf{u}) = \prod_{k=1}^{d}(a_k + 1) \sum_{\mathbf{i} \in \mathbb{Z}^{+d}} p_{\mathbf{i}} \prod_{k=1}^{d} \binom{a_k + i_k + 1}{i_k} u_k^{i_k}(1 - u_k)^{a_k}, \quad \mathbf{u} \in (0,1)^d \qquad (4)$$

with $p_{i_k} = P(Z_k = i_k) = \dfrac{a_k}{(a_k + i_k)(a_k + i_k + 1)}$, $i_k \in \mathbb{Z}^+, k = 1, \cdots, d$.

**Example 2** (Poisson copula): Let, for $L(u) := -\ln(1-u) > 0$, $u \in (0,1)$ and fixed parameters $a_k > 0$,

$$\varphi_{ki}^P(u) = (1-u)^{a_k} \frac{a_k^i L(u)^i}{i!} \tag{5}$$

Here we have $\alpha_{ki}^P = \int_0^1 \varphi_{ki}^P(u)\,du = \left(\frac{a_k}{a_k+1}\right)^i \left(1 - \frac{a_k}{a_k+1}\right)$, representing geometric distributions over $\mathbb{Z}^+$.

The densities $f_{ki}^P$ are those of a transformed gamma distribution with $f_{ki}(u) = (a_k+1)^{i+1} \frac{L^i(u)}{i!}(1-u)^{a_k}$ for $k, i \in \mathbb{Z}^+$ and $u \in (0,1)$. The corresponding copula density is thus given by

$$c^P(\mathbf{u}) = \prod_{k=1}^d (a_k+1) \sum_{\mathbf{i} \in \mathbb{Z}^{+d}} p_{\mathbf{i}} \prod_{k=1}^d \frac{(a_k+1)^k}{i_k!} L^{i_k}(u_k)(1-u_k)^{i_k}, \quad \mathbf{u} \in (0,1)^d \tag{6}$$

with $p_{i_k} = P(Z_k = i_k) = \left(\frac{a_k}{a_k+1}\right)^{i_k}\left(1 - \frac{a_k}{a_k+1}\right) = \frac{a_k^{i_k}}{(a_k+1)^{i_k+1}}$, $i_k \in \mathbb{Z}^+, k = 1,\cdots,d$.

Note that a random variable $X_{ki}$ with density $f_{ki}^P$ can be represented as $X_{ki} = 1 - \exp(-Y_{ki})$, where $Y_{ki}$ follows a gamma distribution with shape parameter $i+1$ and scale parameter $a_k+1$.

In *Pfeifer et al.* (2016), essentially the symmetrical case, i.e. the case of identical components of the driver $\mathbf{Z}$ was considered (so called *diagonal dominance*). This means that the copula induced by the $d$-variate distribution of $\mathbf{Z}$ is given by the upper Fréchet bound $C(\mathbf{u}) = \min(u_1, \cdots, u_d)$, $\mathbf{u} \in (0,1)^d$. In the two-dimensional case, the above formulas simplify to a great extent. In particular, it was proved that the negative binomial copula for arbitrary $a_1 = a_2 = a > 0$ has a positive tail dependence coefficient given by

$$\lambda_U(a) = \lim_{t \uparrow 1} \frac{\int_t^1 \int_t^1 c^{NB}(u,v)\,du\,dv}{1-t} = 1 - \frac{\binom{2a}{a}}{4^a} \sim 1 - \frac{1}{\sqrt{\pi a}} \text{ for large } a. \tag{7}$$

The Poisson copula, in contrast, has no tail dependence for all choices $a_1 = a_2 = a > 0$ although the density given by (6) is unbounded and has a pole in the point $(u,v) = (1,1)$.

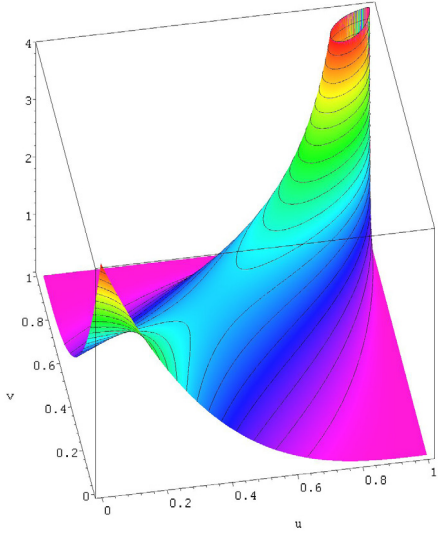 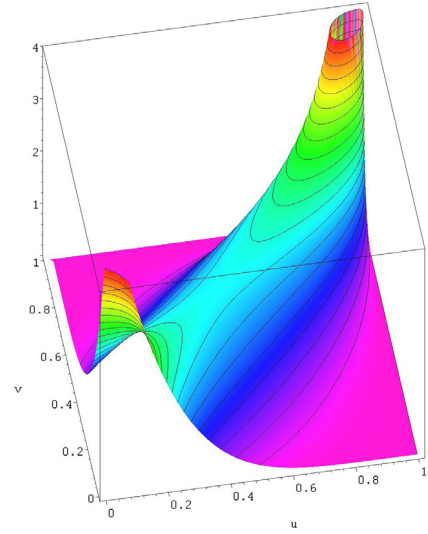

density of a bivariate negative binomial copula, $a = 5$     density of a bivariate Poisson copula, $a = 5$

So far the question remained open how such copulas could be fitted to highly *asymmetric* data in arbitrary dimensions. In the sequel, we shall give a constructive answer to this problem.

## 3. Constructing infinite partition-of-unity copulas from given data

The general idea here is to relate the $\{p_\mathbf{i}\}_{\mathbf{i} \in \mathbb{Z}^{+d}}$, which essentially determine the structure of the IPU copula, to the empirical copula given by the data observed in an appropriate way. Let, for this purpose, denote $\hat{C}$ a copula that is suitably estimated from the empirical copula, like a Bernstein copula, a rook copula (cf. *Cottin and Pfeifer* (2014)), appropriate shuffles of *M* (cf. *Nelsen* (2006), section 3.2.3) or other patchwork copulas that can be easily simulated by Monte Carlo methods (cf. *Durante and Fernández-Sánchez* (2010) or *Durante et al.* (2013)). Let further denote $F_k$ the cumulative distribution function induced by the discrete distribution $\alpha_{k\bullet}$, i.e. $F_k(i) = \sum_{j=0}^{i} \alpha_{kj}$, $i \in \mathbb{Z}^+$. If $\hat{\mathbf{U}} = (\hat{U}_1, \cdots, \hat{U}_d)$ denotes a stochastic representation of $\hat{C}$, set $Z_k := F_k^{-1}(\hat{U}_k)$ for $k = 1, \cdots, d$. Then $\mathbf{Z} = (Z_1, \cdots, Z_d)$ is an appropriate driver for a data-driven IPU copula.

For the examples given above, the resulting formulas are quite simple.

Example 1: $Z_k = F_k^{-1}(\hat{U}_k) = \left\lfloor \dfrac{a_k \hat{U}_k}{1-\hat{U}_k} \right\rfloor$, where $\lfloor z \rfloor = \max\{x \in \mathbb{R} \mid x \leq z\}$ (round down).

Example 2: $Z_k = F_k^{-1}(\hat{U}_k) = \left\lfloor \dfrac{-\ln(1-\hat{U}_k)}{\ln(a_k+1) - \ln a_k} \right\rfloor$.

This follows immediately from standard arguments in Monte Carlo theory: in Example 1, we have
$Z_k = \left\lfloor \dfrac{a_k \hat{U}_k}{1-\hat{U}_k} \right\rfloor = i$ iff $\dfrac{i}{a_k+i} \leq \hat{U}_k < \dfrac{i+1}{a_k+i+1}$, with probabilities $P(Z_k = i) = P\left(\dfrac{i}{a_k+i} \leq \hat{U}_k < \dfrac{i+1}{a_k+i+1}\right)$
$= \dfrac{a_k}{(a_k+i)(a_k+i+1)}$ for $i \in \mathbb{Z}^+$, as desired.

In Example 2, we have $Z_k = F_k^{-1}(\hat{U}_k) = \left\lfloor \dfrac{-\ln(1-\hat{U}_k)}{\ln(a_k+1)-\ln a_k} \right\rfloor = i$ iff $1-\left(\dfrac{a_k}{a_k+1}\right)^i \leq \hat{U}_k < 1-\left(\dfrac{a_k}{a_k+1}\right)^{i+1}$,

with probabilities $P(Z_k = i) = P\left(1-\left(\dfrac{a_k}{a_k+1}\right)^i \leq \hat{U}_k < 1-\left(\dfrac{a_k}{a_k+1}\right)^{i+1}\right) = \dfrac{a_k^i}{(a_k+1)^{i+1}} = \left(\dfrac{a_k}{a_k+1}\right)^i\left(1-\dfrac{a_k}{a_k+1}\right)$

for $i \in \mathbb{Z}^+$, as desired.

The method proposed here allows for a great flexibility concerning the construction of data-driven IPU copulas, including cases with positive tail dependence. We discuss this here along the example given in *Cottin and Pfeifer* (2014), Example 4.2, which was also the basis for the discussion in *Pfeifer et al.* (2016), Section 4.

The following table shows the original data $(x_i, y_i)$ and the corresponding rank vectors $(r_{1i}, r_{2i})$.

| $i$ | $x_i$ | $y_i$ | $r_{1i}$ | $r_{2i}$ |
|---|---|---|---|---|
| 1 | 0.468 | 0.966 | 4 | 9 |
| 2 | 9.951 | 2.679 | 20 | 20 |
| 3 | 0.866 | 0.897 | 8 | 4 |
| 4 | 6.731 | 2.249 | 19 | 19 |
| 5 | 1.421 | 0.956 | 13 | 8 |
| 6 | 2.040 | 1.141 | 17 | 15 |
| 7 | 2.967 | 1.707 | 18 | 18 |
| 8 | 1.200 | 1.008 | 11 | 10 |
| 9 | 0.426 | 1.065 | 3 | 12 |
| 10 | 1.946 | 1.162 | 15 | 16 |
| 11 | 0.676 | 0.918 | 5 | 6 |
| 12 | 1.184 | 1.336 | 10 | 17 |
| 13 | 0.960 | 0.933 | 9 | 7 |
| 14 | 1.972 | 1.077 | 16 | 13 |
| 15 | 1.549 | 1.041 | 14 | 11 |
| 16 | 0.819 | 0.899 | 6 | 5 |
| 17 | 0.063 | 0.710 | 1 | 1 |
| 18 | 1.280 | 1.118 | 12 | 14 |
| 19 | 0.824 | 0.894 | 7 | 3 |
| 20 | 0.227 | 0.837 | 2 | 2 |

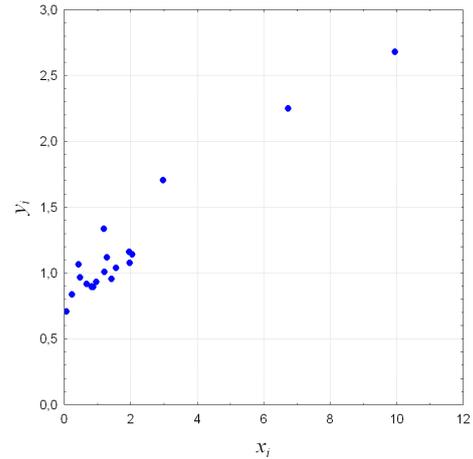

graph of original data

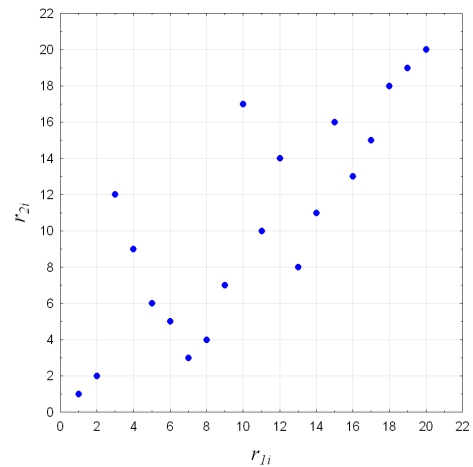

graph of rank vectors

The following graphs show 5,000 Monte Carlo simulations each from different constructions of data-driven IPU copulas (small dots), with a superposition of the empirical copula (scaled rank vectors) as large white points. The symmetric cases (negative binomial and Poisson copulas) were constructed according to the suggestions in *Pfeifer et al.* (2016), Section 4. The asymmetric cases were constructed on the basis of a shuffle of *M* copula with local upper Fréchet bounds for the driver **Z**, shown first. Note that the location of the corresponding line sections are one to one determined by the relative rank vectors from the original data.

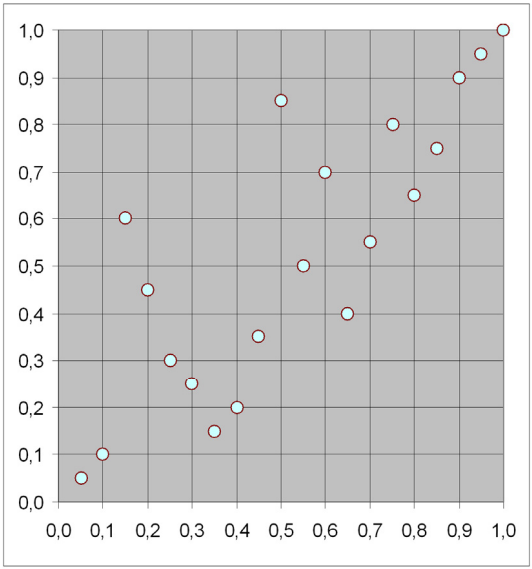

relative ranks $r_{\bullet i}/20$

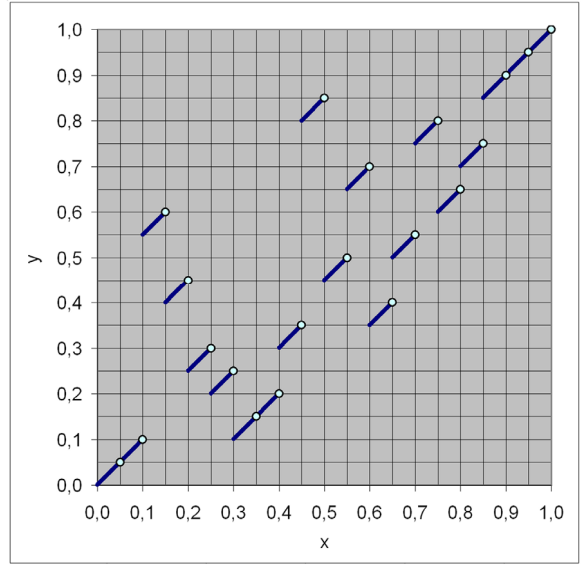

shuffle of *M* with local upper Fréchet bounds

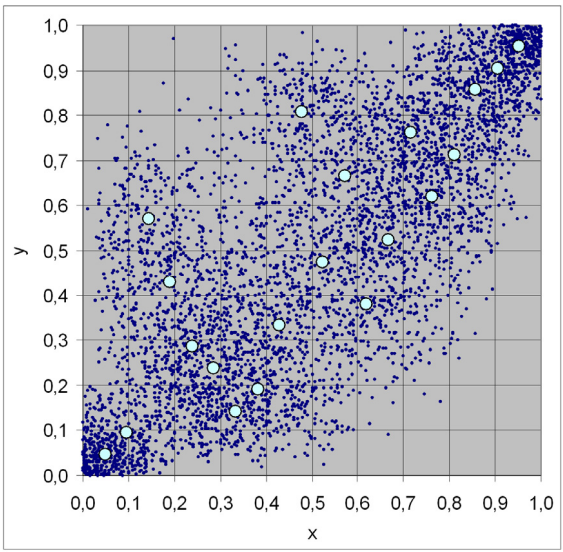

Bernstein copula

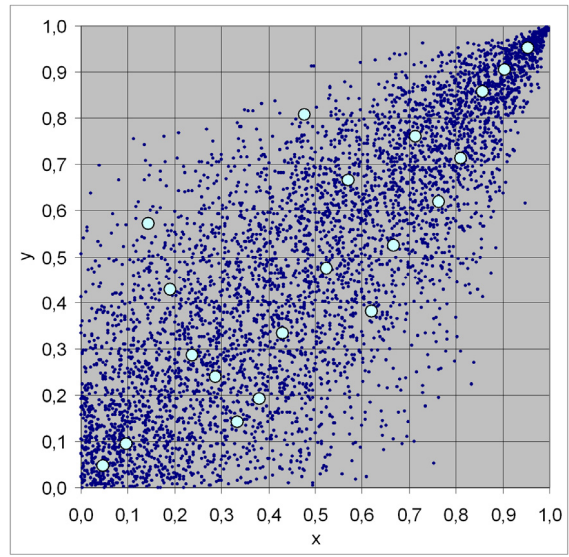

symmetric negative binomial copula, $a = b = 5$

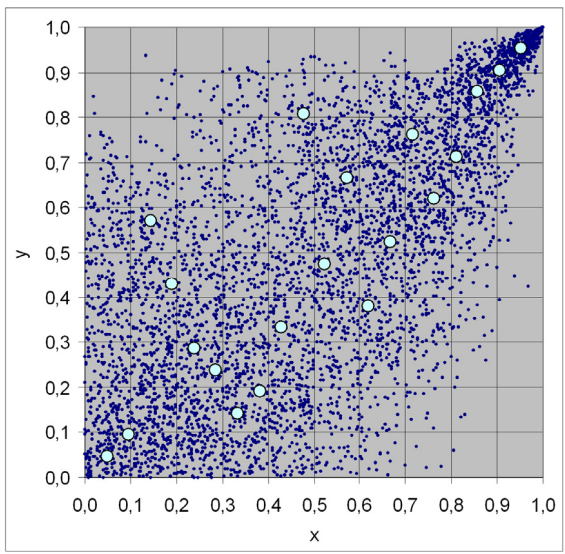
asymmetric negative binomial copula, $a = b = 5$

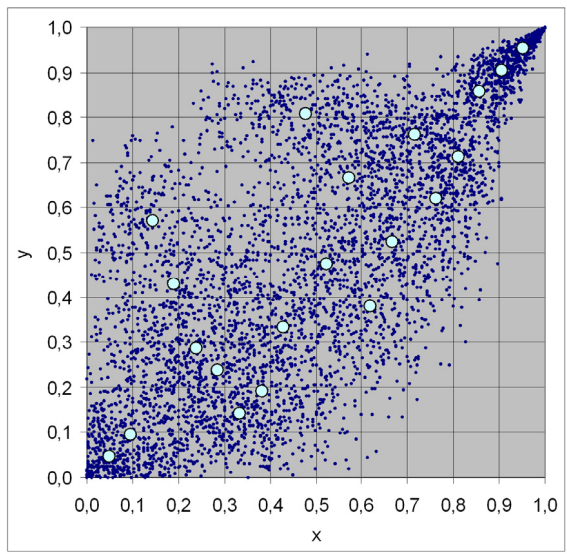
asymmetric negative binomial copula, $a = b = 10$

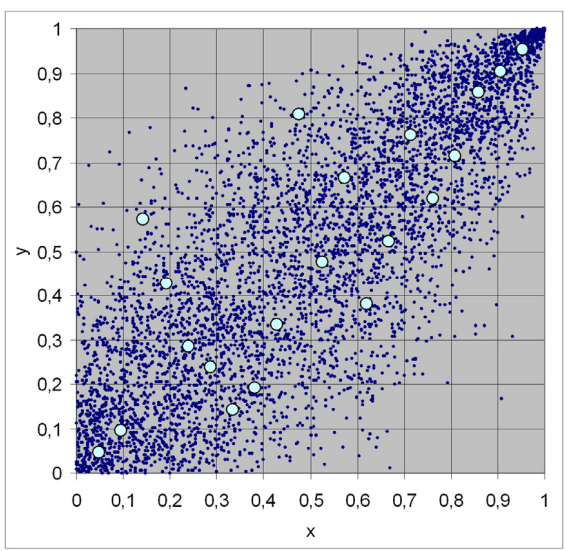
symmetric Poisson copula, $a = b = 6$

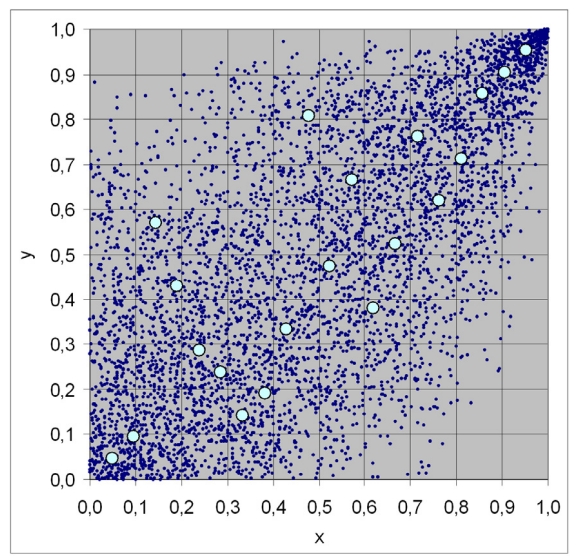
asymmetric Poisson copula, $a = b = 6$

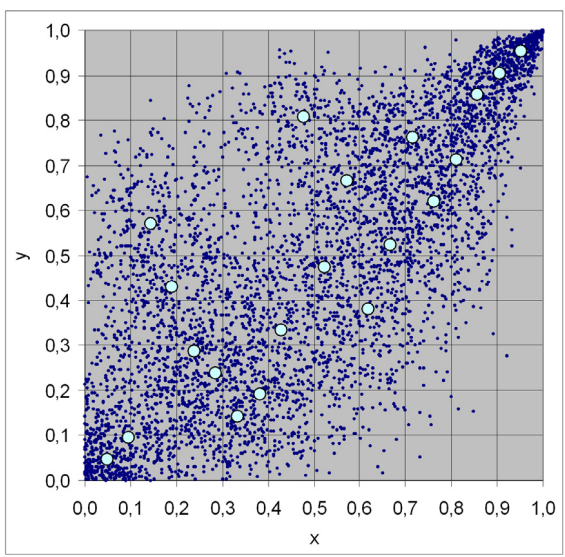
asymmetric Poisson copula, $a = b = 10$

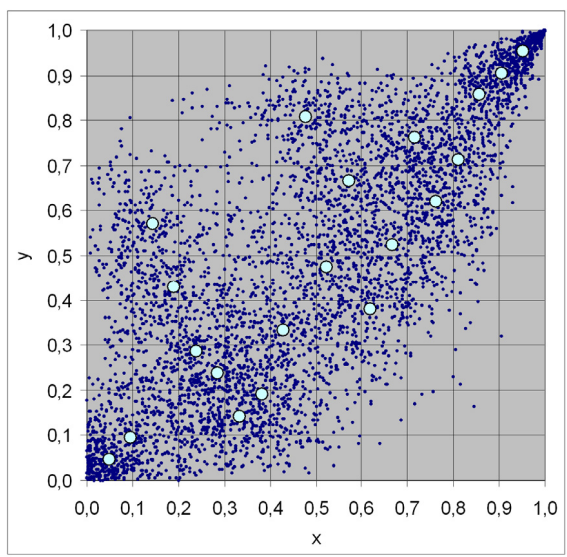
asymmetric Poisson copula, $a = b = 15$

As can be clearly seen, the asymmetric IPU copula follows the given data (empirical copula) much better than the symmetric IPU copulas. Also, in contrast to the Bernstein copula with no tail dependence, the asymmetric negative binomial IPU copulas always show a positive tail dependence. Since tail dependence is an asymptotic property, we can conclude that the corresponding tail dependence coefficient in the above constructions can be calculated from the parameter $a=b$ according to relation (7). Note also that although the pictures above might suggest a positive tail dependence for the asymmetric Poisson IPU copula, this is theoretically not possible.

## 4. Implications for risk management

The new European supervisory regulations in the financial sector (Basel III for banks, Solvency II for insurance companies) require the calculation of a sufficient capital adequacy based on the risk measure Value@Risk $\text{VaR}_\alpha(S)$, which is defined as the $1-\alpha$ quantile of the distribution of the total portfolio risk $S = \sum_{i=1}^{d} X_i$ where $X_1, \cdots, X_d$ are the individual risk positions in the portfolio. Especially in internal models for the calculation of the underlying aggregate risk measure, it is important to find appropriate models for the stochastic dependence between individual risk positions. It is well known that in the case of comonotonicity between risks – i.e., the underlying copula is the upper Fréchet bound – there is no diversification effect and the risk measure Value@Risk is additive (cf. *McNeil et al.* (2015), Proposition 7.20). Also, the worst case for the total Value@Risk is not attained under comonotonicity but rather in cases where there is a kind of local negative dependence in the upper right corner of the underlying copula (cf. *Puccetti and Rüschendorf* (2012) or *Embrechts et al.* (2013)). A similar negative result holds for an assumed dependence between correlation and diversification (cf. *Pfeifer* (2013)), which is frequently stated in the common legislative papers. The following examples show how the asymmetric data-driven IPU copula approach can provide competing estimates for the risk measure Value@Risk on the basis of the same data observed. For the sake of simplicity, we use the data set from *Cottin and Pfeifer* (2014), Example 4.2, discussed above. We compare the following IPU copula approaches:

- a classical Bernstein copula with grid size 20 (cf. *Cottin and Pfeifer* (2014))
- the asymmetric negative binomial copula with parameters $a=b=5$, $a=b=10$ and $a=b=15$ (for short: NB5, NB 10 and NB 15)
- the asymmetric Poisson copula with parameters $a=b=6$, $a=b=10$ and $a=b=15$ (for short: Po 6, Po 10 and Po15)
- "worst case" (WC) versions of these copulas where a particular shuffle of *M* is used (i.e. with a local lower Fréchet bound in the upper right corner)

The following graphs show the corresponding "worst case" copula driver:

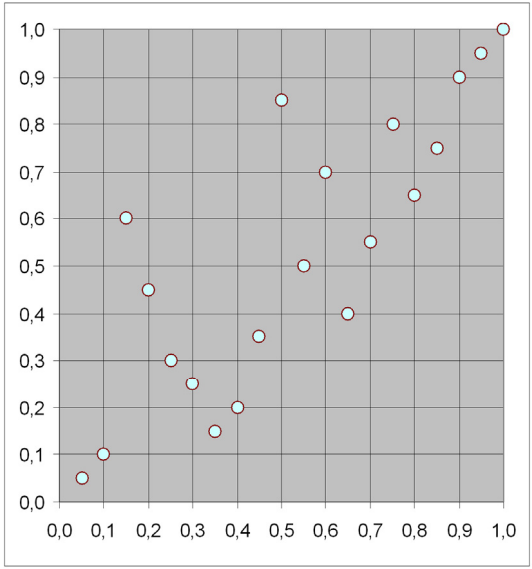
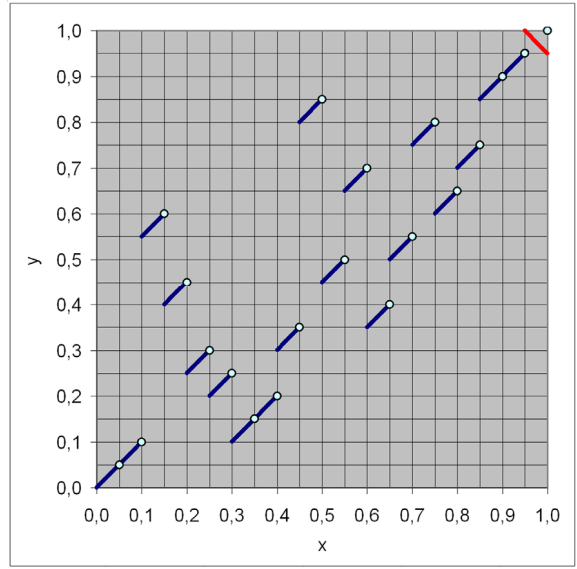

relative ranks $r_{\bullet i} / 20$          "worst case" shuffle of $M$

The following graphs show scatterplots from 5,000 simulations each for the underlying "worst case" copulas:

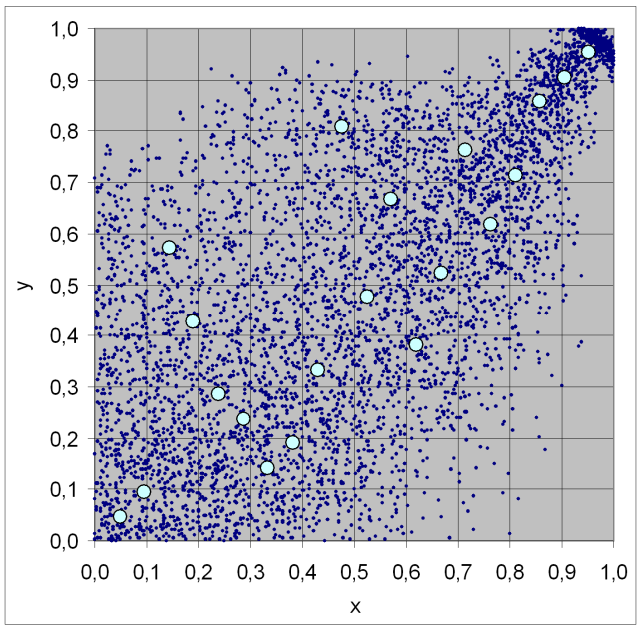
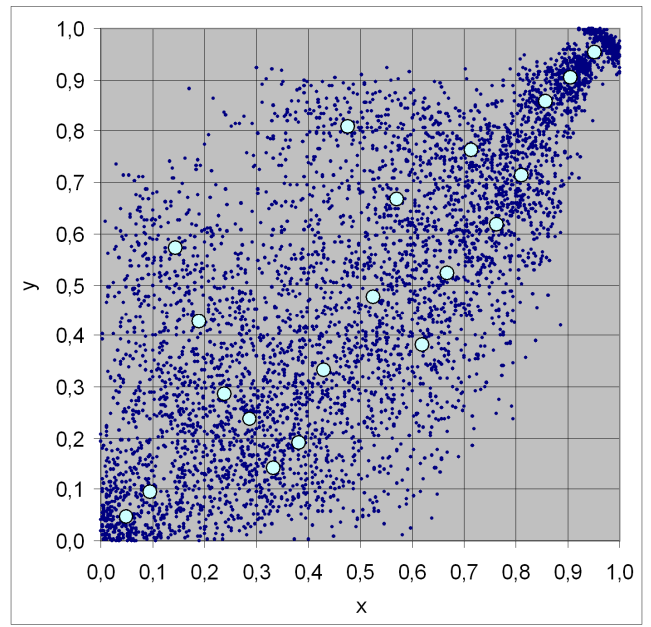

"worst case" negative binomial copula, $a = b = 5$     "worst case" negative binomial copula, $a = b = 10$

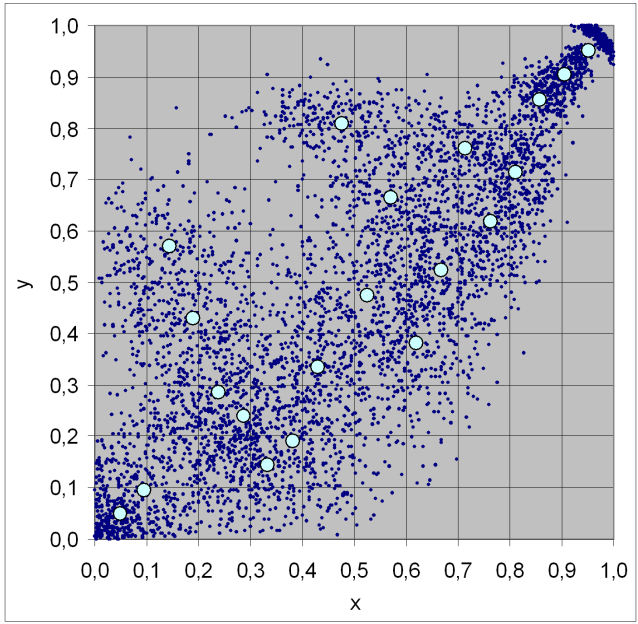

"worst case" negative binomial copula, $a=b=15$

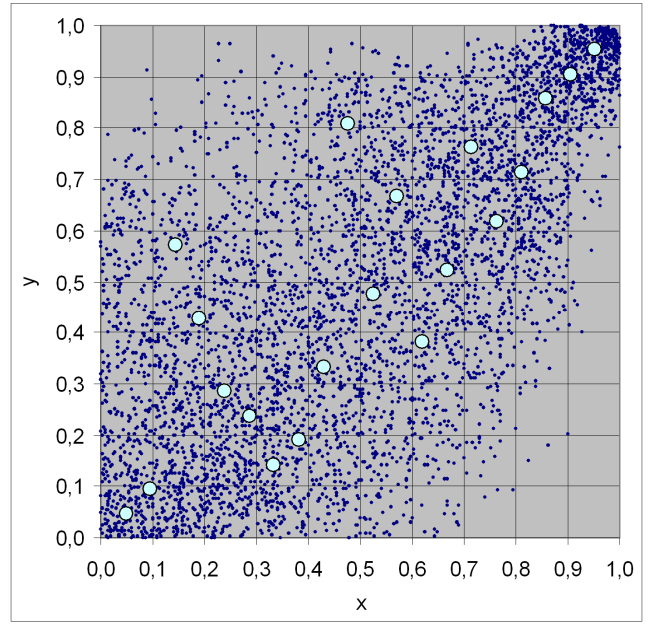

"worst case" Poisson copula, $a=b=6$

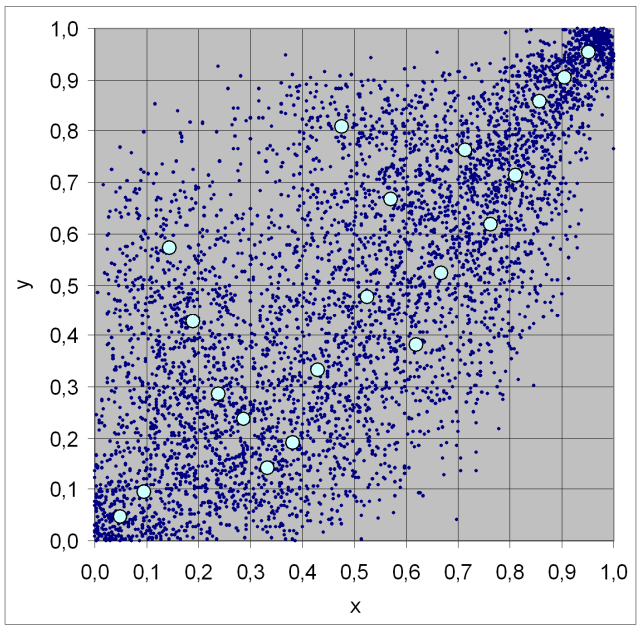

"worst case" Poisson copula, $a=b=10$

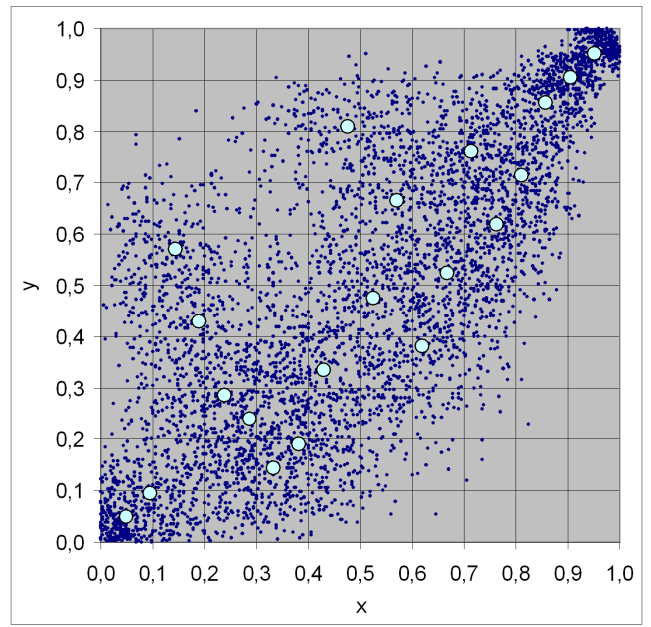

"worst case" Poisson copula, $a=b=15$

The following estimates are based on 5,000,000 Monte Carlo simulations for each particular copula approach. For the marginal distributions of the two risk positions $X$ and $Y$, a lognormal and a Fréchet distribution were estimated from the original data. The risk level $\alpha$ was chosen as $\alpha = 0.05$.

With the estimated parameters for the marginal distributions, we have $\widehat{\text{VaR}}_{0,05}(X) = 6.8190$, $\widehat{\text{VaR}}_{0,05}(Y) = 2.0984$ and $\widehat{\text{VaR}}_{0,05}(X) + \widehat{\text{VaR}}_{0,05}(Y) = 8.9174$. From the simulations, we obtain, with $S = X + Y$,

| Copula type | Bernstein | NB 5 | NB 5 WC | NB 10 | NB 10 WC | NB 15 | NB 15 WC |
|---|---|---|---|---|---|---|---|
| $\widehat{\text{VaR}}_{0.05}(S)$ | **8.9586** | **8.8474** | **9.3989** | **8.8834** | **9.5421** | **8.8978** | **9.6198** |
| $\widehat{\text{VaR}}_{0.05}(X)+\widehat{\text{VaR}}_{0.05}(Y)$ | 8.9174 | 8.9174 | 8.9174 | 8.9174 | 8.9174 | 8.9174 | 8.9174 |

| Copula type | Po 6 | Po 6 WC | Po 10 | Po 10 WC | Po 15 | Po 15 WC |
|---|---|---|---|---|---|---|
| $\widehat{\text{VaR}}_{0.05}(S)$ | **8.8200** | **9.1402** | **8.8453** | **9.2412** | **8.8820** | **9.3532** |
| $\widehat{\text{VaR}}_{0.05}(X)+\widehat{\text{VaR}}_{0.05}(Y)$ | 8.9174 | 8.9174 | 8.9174 | 8.9174 | 8.9174 | 8.9174 |

The following graphs show some empirical quantile functions of the simulations above:

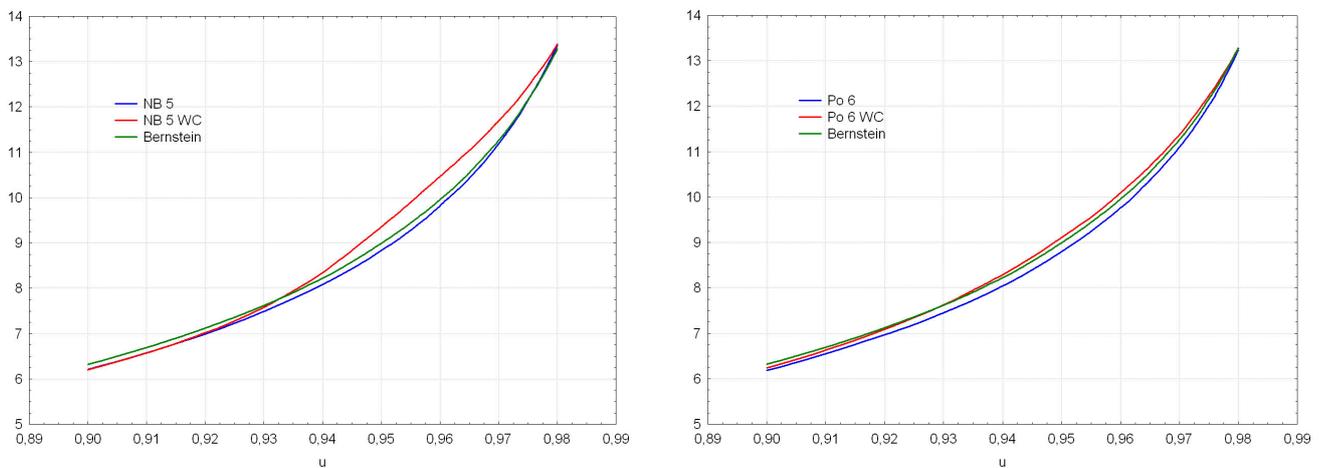

empirical quantile functions

These results clearly show that the "worst case" IPU copula approaches always result in a risk concentration effect while the basic negative binomial and the Poisson copula approach show a slight diversification effect which decreases with increasing parameters $a=b$. Note that the results for the negative binomial and the Poisson copula are quite close in spite of the fact that the negative binomial copula here always has a positive tail dependence. It is interesting to see that the Bernstein copula also shows a risk concentration effect although there is no tail dependence and also no strict "worst case" behaviour.

## 5. Conclusion

The IPU copula approach for asymmetric data sets is a very flexible tool to model dependencies between risks, also in higher dimensions. It covers cases of tail dependence and also of "worst case" scenarios on the basis of the same data set. It follows the shape of the data more closely than most other approaches and can easily be implemented in usual spreadsheets. Note that our motivation for a patchwork construction for the copula driver resembles very much the arguments in *Durante et al.*(2013). The difference is, however, that the resulting IPU copula itself is not a patchwork copula.

Especially in the light of the new European supervisory regulations in the financial sector such approaches might be interesting to figure out unfavourable constellations which lead to a higher demand of equity or solvency capital. It should be kept in mind, however, that this is not only a problem of the as-sumed underlying copula, but also depends significantly on the type of the marginal risk distributions, as is discussed in *Ibragimov and Prohorov* (2017).

**Acknowledgement:** We would like to thank the referees for some critical comments on an earlier version which lead to a clearer exposition of the content of the paper.